\documentclass[12pt]{article}
\usepackage[dvips]{color,graphicx}
\pagecolor{white}
\def\bfq {{\bf q}}
\def\bfP{{\bf P}}
\def\bfQ{{\bf Q}}
\def\bfB{{\bf B}}\def\bfk{{\bf k}}
  
\def\bfb{{\bf b}}
\newcommand{\bfkappa}{\mbox{\boldmath $\kappa$}}
  \def\be{\begin{equation}}
 \def \ee{\end{equation}}
\def\bea{\begin{eqnarray}}
  \def\eea{\end{eqnarray}}

\begin{document}

\title{\hskip12.0cm {\small NT@UW-03-035}
\\Color Transparent GPDs?}

\author{Matthias Burkardt\\
Department of Physics,
New Mexico State University\\
Las Cruces, NM 88003-0001\\
e-mail: burkardt@nmsu.edu\\
 Gerald A. Miller\\
Department of Physics, University of Washington\\
 Seattle, WA 98195-1560\\
e-mail: miller@phys.washington.edu}

\maketitle

\abstract{The relation between GPD's and color transparency is explored.
The discovery of color transparency in pionic 
diffractive dissociation reactions
allows us to make specific predictions for the behavior of 
the pion generalized parton distribution, and provide a further test of
 any  model of the   pion form factor.}

\section{Introduction}
Color transparency is the vanishing of initial and final state interactions,
predicted by
QCD to occur in coherent nuclear processes for large values of  transferred
momentum\cite{mueller,brodsky,Frankfurt:1992dx,FMS94,Jain:1995dd}. 
For such processes, the interactions 
of a color singlet object 
are controlled by the  color electric dipole moment which is small 
if the quark-gluon constituents 
          are close together.
  
Consider, for example, the $(e,e^\prime p)$ reaction for quasi-elastic
kinematics.
Color transparency may occur when those components of the hadron
wave function that dominate the proton form factor at large $Q^2$  have a
small transverse size. This is indeed the case if 
the form factor at large momentum transfer is correctly described
by perturbative Quantum Chromodynamics, pQCD. 
However, the power law behavior obtained
from pQCD arguments can also be obtained from the Feynman mechanism
in which  the active quark carries 
almost all the $p^+$ momentum of the hadron. The spectator quarks 
are then ``wee partons'' which have no direction and hence it is
very easy to turn them around. The wee parton cloud is  not  
expected  to be of small  transverse size  and therefore
such components of the wave function would interact strongly as the
emitted object moves 
 out of the nucleus. Then 
color transparency would not be observed.
 Color transparency is not a necessary result  
 even if a small-sized configuration
dominates the form factor\cite{FMS94}. 
The configuration must remain of small size during
the time required for escape from the nucleus, and this requires high energies.

Dominance of small-size, or point-like configurations, 
at a high momentum transfers is
a necessary condition on the existence 
of color transparency, and it is  worthwhile 
to see if this can be related to other features of hadronic wave functions.
Here we shall examine generalized parton distributions (GPDs) which have the
 amazing property  of
 providing a decomposition of the form factor with
respect to the light-cone momentum of the active quark
\bea
F_1(Q^2) = \int dx H(x,0,Q^2).
\label{eq:FH}
\eea
Eq. (\ref{eq:FH}) implies that if, we know the function
$H(x,0,Q^2)$ at large $Q^2$, then we also know the important 
regions of $x$ for the form factor. This knowledge should enable us
to differentiate between suggested mechanisms that drive form factors
at large $Q^2$ and should
ultimately   allow us to understand whether or not color 
transparency may  occur for the proton.  

%%new
Providing a logical link between GPDs and color transparency would
allow physicists to answer some general questions.
Can the observation of color transparency determine any features of GPDs?
Conversely, could  a measurement  of  a GPD for a given  system enable us
to predict the ability to observe the effects of color transparency?

To answer these questions it is necessary to review some background 
information about color transparency, CT. There are three requirements for
 CT to occur\cite{Jennings:1989hc} in coherent reactions:
\begin{itemize}
\item
 small color neutral objects do not interact

\item  high momentum transfer (semi) exclusive reactions proceed through
configurations
of small size, that we call  point like configurations, PLC. This
depends on the process occurring with 
a high enough momentum transfer. 

\item  the PLC must move quickly enough through the nucleus to escape prior
to expansion (which is inevitable). This depends on the PLC having large
energy 
since the time dilation factor is the energy divided by a mass.
\end{itemize}

The second requirement addresses the most interesting aspect of
 color transparency physics: the necessity of the formation of a PLC for a 
 high momentum transfer coherent reaction to  proceed, and is
our focus.
 The formation of a PLC  is a natural occurrence in PQCD\cite{mueller,brodsky},
but may (and perhaps can be expected to)
arise in strong QCD\cite{Frankfurt:1992dx,prec}.

 An operational
 method to  test whether
or not a given model for a hadron allows the formation of 
PLC was devised\cite{Frankfurt:1992dx,prec}. Let's use a schematic 
notation to understand the basic idea. Suppose one has an initial hadron 
$\vert H\rangle$ that is subject to a high momentum transfer reaction. Let the
hard operator that brings in the high momentum be denoted as $T_H(Q^2)$.
The operator $T_H(Q^2)$ arises solely from the gluon exchanges within the
hadron $H$, and depend only on the nature of the target.
 The 
resulting wave packet is denoted as 
$T_H(Q^2)\vert H\rangle$, which or may or may not be a $\vert{\rm PLC}\rangle$.
 Suppose 
this object propagates through the nucleus, and is detected, moving with 
the transferred momentum. Then the scattering amplitude is given by 
\bea {\cal M}=\langle H(Q^2)\vert \left(1+ T_SG_0  
\right)T_H(Q^2)\vert H\rangle,
\label{mult}\eea
where the operator $G_0$ represents the free propagation and 
$T_S$ denotes the soft final state interaction, to all orders of interaction
($T_S=U_S(1+G_0T_S))$,
 with the nucleus. 
If $T_H(Q^2)\vert H\rangle$ is a PLC (which is a wave packet and not an
eigenstate of the Hamiltonian) 
the inevitable action of the propagation
is to cause the object to expand. Suppose the energy is large enough for 
the packet to remain in a fixed configuration as it moves through the nucleus.
Then the influence of the final state interactions is determined by 
$\langle H(Q^2)\vert U_ST_H(Q^2)\vert H\rangle$.
 If the struck object really is 
small, the scattering amplitude for the interaction of an energetic, 
colorless wave packet of small transverse size, characterized by a length $b$,
is  proportional to $b^2$ 
(times logarithmic corrections)\cite{Low:1975sv}-\cite{Frankfurt:1996ri}.
This dependence  can be thought qualitatively as arising
 from  the action of two color dipole operators. Then
\bea U_S\propto b^2,
\label{usb2}\eea and 
the relevant matrix element that determines the importance of final state 
interactions is
 $\langle H(Q^2)\vert b^2T_H(Q^2)\vert H\rangle$, and the strength of
the second term of Eqn.~(\ref{mult}) compared to the  first term is determined
by the ratio 
$\langle H(Q^2)\vert b^2T_H(Q^2)
\vert H\rangle/\langle H(Q^2)\vert T_H(Q^2)\vert H\rangle$.

 The denominator is simply the hadronic form factor, $F(Q^2)$, so that we 
may  define an  effective size $b^2(Q^2)$:
\bea
b^2(Q^2)={\langle H(Q^2)\vert b^2T_H(Q^2)\vert H\rangle\over F(Q^2)}
 .\label{b22} \eea
CT can only occur if $b^2(Q^2)$  is much  smaller than the mean square 
radius  of the hadron. A large magnitude of  $b^2(Q^2)$ causes
 the final state interactions to  be large so that  
 CT would not occur. 
%%here is where positivity is addressed
Note that $b^2(Q^2)$ depends on the
full eigenstate, and as an-off diagonal
 matrix element, is not positive definite. 
The term $b^2(Q^2)$ takes its meaning as an effective size 
from its operational importance in determining whether
or not final state interactions will occur.

We want to use the important fact that
 color transparency has been observed in diffractive 
dissociation of high energy pions\cite{Aitala:2000hb,Aitala:2000hc}.
The cross sections were found to  
scale roughly as $A^{\alpha}$ (with $\alpha$ varying between 4/3 and 5/3)
 instead of the usual
 $A^{2/3}$ dependence typical of diffractive cross sections. The 
consequence is that CT causes 
 the ratio  of the cross sections for Platinum to Carbon targets 
to increase by  a spectacular factor
of 7\cite{Frankfurt:2000jm}. Indeed, 
the relative simplicity of the pionic wave function, the availability of
a high energy pion beam, and predicted 
unusual $A$-dependence \cite{Frankfurt:it} made
this reaction ideal for studies of color transparency.

Here is an outline of the remainder of this paper. The relation between
color transparency and GPDs is explored 
in Sect. 2  by considering the effective
size\cite{Frankfurt:1992dx,prec} of  an emitted particle  
using an impact representation\cite{mb1}.
That the di-jet measurements are shown to imply a very small 
value of 
$b^2(Q^2)$ is shown in Sect. ~3. As a result, most
 of the remainder is concerned
with the pion.  
 The effective size is computed
for a variety of models of GPDs, ranging from very simple factorized forms
to more complicated wave function models that include the effects of
quark spin, in Sect.~4. These studies allow us to arrive at precise 
statements between GPDs and the existence of color transparency that are
summarized in Sect.~5. If we use the experimental 
observation of color transparency\cite{Aitala:2000hb,Aitala:2000hc}, then
some models may be ruled out, and the behavior 
of the GPD at large momentum fraction $x$ is determined.

\section{Master Formula}

Since CT is related  to  configurations that are small
in position space, it is very useful to start from an impact
parameter description \cite{mb1}. The distribution of partons in
impact parameter space can be easily obtained from generalized
parton distributions by means of a Fourier transform, yielding
\bea
{\cal H}(x,{\bf B}) = \int \frac{d^2Q}{(2\pi)^2} 
e^{i{\bf B}\cdot {\bf Q}}H(x,0,Q^2).
\label{scriptH}
\eea
Together with Eq. (\ref{eq:FH}) this provides an impact parameter
space representation for the form factor \cite{soper}
\bea F(Q^2)=\int dx H(x,0,Q^2) =
\int dx d^2B \;{\cal H}(x,{\bf B}) 
e^{-i{\bf B}\cdot {\bf Q}}
\label{eq:ff}
\eea
The impact parameter ${\bf B}$ is measured relative to the
transverse center of momentum of the hadron. This impact parameter
can be easily related to the variable ${\bf b}$ which measures the
distance between the active quark and the center of momentum of all
the spectators
\bea
{\bf B}=(1-x){\bf b}.
\eea
At large $Q^2$ one expects that the form factor is expected to be
dominated by the valence component of the wave function.
Therefore in the case of meson form factors
${\bf b}={\bf B}/(1-x)$ is a direct measure of the size of the
system. In a nucleon (and also in a non-valence configuration of a 
meson), knowledge of ${\bf b}={\bf B}/(1-x)$ still
provides a lower bound on the overall size of the system.
It is thus useful to define a $Q^2$-dependent size $b^2(Q^2)$.
The interaction of a struck nucleon with the surrounding medium
depends on the size in a direction transverse to that of the large
momentum ${\bf Q}$. Consequently we take ${\bf Q}$ to lie in the
$x$-direction and study the size of the $y$ component of ${\bf B}$.
Under the condition that ${\cal H}(x,{\bf B}) ={\cal H}(x,{\bf B}\cdot{\bf B})=
{\cal H}(x,B)$ it is straightforward to obtain\cite{prec}
\bea
b^2(Q^2)F(Q^2)&=&{1\over2}
\int dx d^2B\;{{ B_y}^2\over (1-x)^2} 
{\cal H}(x,{\bf B}) e^{-i{ B_x} { Q}}\nonumber\\
&=&-\left({\partial \over\partial  Q^2}\right)
\int dx d^2B\;{1 \over(1-x)^2} {\cal H}(x,{\bf B}) 
e^{-i{\bf B}\cdot {\bf Q}}.
%\nonumber\\
%&=&-\left({\partial^2 \over\partial  Q^2}+
%{1\over Q} {\partial \over \partial Q}\right)
%\int \frac{dx}{(1-x)^2} H(x,0,Q^2).
\label{eq:master0}
\eea
Using Eq. (\ref{eq:ff}), this implies an interesting
relation between GPDs and the effective size of
a hadron at large $Q^2$
\bea
b^2(Q^2)= -\frac{\left({\partial \over\partial  Q^2}\right)
\int \frac{dx}{(1-x)^2} H(x,0,Q^2)
}{\int_0^1 dx H(x,0,Q^2)}.
\label{eq:master}
\eea
The rest of the paper will be devoted to
discussing the implications of this result for the
behavior of GPDs.

Notice the factor $\frac{1}{(1-x)^2}$ in Eq. (\ref{eq:master}), which
appears because the distance ${\bf b}$ 
between the active quark and the
spectator(s) is larger than the impact parameter
${\bf B}$ (the distance to the center of momentum) by a factor
$1/(1-x)$. This factor diverges for $x\rightarrow 1$, allowing
a hadron of ordinary size  to support a large
form factor if the integrand of Eq.~(\ref{eq:ff}) is 
dominated contributions from the region with
momentum fraction $x$ near $1$. 
In particular, 
the main issue is that the variable conjugate to $Q$ is the impact 
parameter which needs to be divided by $1-x$ in order to obtain the
separation between the active quark and the spectator(s) --- which is a better
measure for the ``size'' of the hadron than the impact parameter.

A necessary condition for color transparency to occur is
that \bea \lim_{Q\to\infty}b^2(Q^2)=0.
\label{necessary}
\eea
This condition is not sufficient because the distance from the active
quark to the center of momentum of the spectators only provides 
a lower 
bound on the size of the configuration if there is more than one
spectator parton. Furthermore $\int d^2B B_y^2 {\cal H}(x,{\bf B})
e^{-iB_xQ}$ is strictly speaking not positive definite, although it
turns out to be positive for commonly used parametrizations of
GPDs at large $Q$ --- at least for those where ${\cal H}(x,{\bf B})$
satisfies the usual positivity constraints \cite{pos}.

Nevertheless, since Eq. (\ref{necessary}) is a necessary condition
for small since configurations, the existence of color transparency 
would constrain the possible analytic behavior of GPDs as we will 
discuss in the rest of the paper.

\section {Color Transparency in Di-jet Production and $b^2(Q^2)$}
To utilize the experimental discovery, 
it is necessary to relate di-jet production
to  color transparency in
the $(e,e' \pi)$ reaction, which  is closely related to the
physics of the
pion  form factor discussed in Eqs.~(\ref{mult}-\ref{b22}). In this
connection, it is useful to explicitly present  the 
 coordinate-space representation for the putative PLC:
\bea
\langle x,\bfb\vert T_H(Q^2)\vert\pi\rangle=
e^{i\bfq\cdot\bfb(1-x)}\psi_\pi(x,\bfb)=\langle x,\bfb \vert{\rm PLC}(Q^2)\rangle.
 \label{coord}
\eea
In coherent nuclear 
 di-jet production the  incident pion beam 
is  changed into a  state consisting of a $q,\bar{q}$ pair moving at high
relative transverse momentum (${\bf \kappa}$, 
but with a total momentum close to $\bfP_\pi$. The quark has longitudinal momentum $x\bfP_\pi$, and the cross section is dominated by the region 
$x\approx1/2 .$
Each quark becomes a hadronic jet at distances outside the target.

The interactions 
with the nuclear target must be of low momentum transfer because
 the target is
not excited, so that 
  the di-jet cross section 
depends on the matrix element\cite{Frankfurt:it}
\bea {\cal M}^{\rm JJ}=\langle x, \bfkappa\vert \left(1+ T_SG_0  
\right)U_S(Q^2)\vert \pi\rangle,
\label{mult1}\eea
Where ${\bfkappa}= (1-x){\bf k_1} - x {\bf k_2}$ and
${\bf k_1}, {\bf k_2}$ represent the momenta of $q\bar{q}$ pair
 produced at high relative momentum. Each of these objects   produces 
a jet. For a coherent process to occur, 
the sum of their longitudinal momenta must be close to the
 pion beam momentum, the sum of the transverse momenta must be very small.

The observation of color transparency is tells us that the effects
of final and initial
state interactions are negligible, so that the ratio
\bea
b^2_{\rm JJ}(Q^2)={\langle x,\bfkappa\vert b^2U_S(Q^2)\vert \pi\rangle\over
\langle x, \bfkappa\vert  U_S(Q^2)\vert \pi\rangle}=
{\langle x,\bfkappa\vert b^4\vert \pi\rangle\over
\langle x, \bfkappa\vert  b^2\vert \pi\rangle},
  \eea
in which the second equation is obtained using Eq.~(\ref{usb2}),
must be much smaller in magnitude than the mean square radius of the pion.
 If this ratio were not small, the effects of
initial and final state interactions would cause the $A$ dependence to
differ drastically from what was observed.

The small nature of $b^2_{\rm JJ}(Q^2)$ means that 
$\langle x,\bfkappa\vert U_S\vert \pi\rangle$ acts as a PLC, and that
\bea {\cal M}^{\rm JJ}\approx\langle x,\bfkappa\vert U_S\vert \pi\rangle
\propto\int d^2b e^{-i\bfkappa\cdot\bfb}b^2\psi_\pi(x,\bfb)
=\int d^2b b^2\langle x,\bfb\vert {\rm PLC}(Q^2)
\rangle,\label{mjj}\eea
where here \bea Q^2=
({\bfkappa\over 1-x})^2\approx 4\bfkappa^2.\label{kin}\eea
 Eqn.~(\ref{mjj})  shows  the strong connection between the physics 
of the form factor (which is the integral of Eqn. (\ref{mjj})
without the factor of $b^2$)
and that of di-jet production. We may integrate Eqn.~(\ref{mjj}) by parts to obtain
\bea
 {\cal M}^{\rm JJ}\propto\nabla_\kappa^2\int d^2b
 \langle x,\bfb\vert {\rm PLC}(Q^2)\rangle
\label{mult3}\\
\propto {1\over \kappa^2} \int d^2b  \langle x,\bfb\vert {\rm PLC}(Q^2)\rangle
,\label{mult4}
\eea 
in which the power law falloff of the cross section with $\kappa^2$ was 
used to obtain the result (\ref{mult4}).  
 Eqs.~(\ref{coord}) and (\ref{mult4}) tell us that 
 because  $\vert {\rm PLC}(Q^2)\rangle$ acts as 
small in the di-jet production reaction, $b^2(Q^2)$ must be small,
 for the kinematics of (\ref{kin}). We wish to explore the 
consequences of that smallness for pionic GPDs, and as a result
 much
of this paper is concerned with pionic models. 

\section{The pion form factor}
Empirically, it has been established that the pion form factor
falls like
\bea
F(Q^2) = \frac{c}{Q^2}
\eea
at large $Q^2$ (modulo logarithmic corrections). 
There are infinitely many different possible
ans\"atze for $H(x,0,Q^2)$ that are consistent with 
this behavior and therefore we need to restrict 
ourselves to those classes of functions that are 
most commonly used to parameterize $H(x,0,Q^2)$. 

\subsection{Factorisable Models}
The first class of models that we consider are such that the large 
$Q^2$ behavior factorizes from the $x$-dependence
\be
H(x,0,Q^2) \sim f(x) F(Q^2) \quad \quad \mbox{for}
\quad Q^2\rightarrow \infty .
\label{eq:factor}
\ee
This class covers the whole set of models, where  
$H(x,0,Q^2)$ already has a $1/Q^2$ behavior.
An example is provided by the ``asymptotic'' pion wave function,
which yields
\be
H(x,0,Q^2) \sim \frac{x^2(1-x)^2}{Q^2} \quad \quad \mbox{for}
\quad Q^2\rightarrow \infty .
\ee

Clearly, $b^2(Q^2) \sim {1\over Q^2}$ for all these examples,
provided $\int dx\frac{f(x)}{(1-x)^2}$ converges.
The pathetic case, where $\int dx\frac{f(x)}{(1-x)^2}$ diverges,
corresponds to a hadron that has an infinite size, and we
therefore do not discuss  this possibility. 
As a result, we find color transparency for all GPDs, where the
large $Q^2$ behavior factorizes (\ref{eq:factor}).

\subsection{Exponential Models}
Another important class of models is the one where $H(x,0,Q^2)$
at fixed $x$ falls faster than
$1/Q^2$, e.g. a Gaussian model\cite{Radyushkin:1998rt} where
\bea
H(x,0,Q^2) = f(x) \exp \left( -a^2 Q^2 \frac{1-x}{x}\right).
\eea
In this class of models, obtaining the
 $Q^2\rightarrow \infty$ behavior of both $F(Q^2)$ and
$b^2(Q^2)$, depends on  the crucial region in the 
 vicinity of $x=1$. This  is why all
factors of $x$ are irrelevant for the discussion and we drop
them for simplicity, by considering the class of functions
\bea
H(x,0,Q^2) \stackrel{x\rightarrow 1}
{\longrightarrow} (1-x)^{m-1} \exp \left( -a(1-x)^nQ^2\right).
\label{eq:hx}
\eea
The common feature of  these models is that the $Q^2$ dependence
disappears as $x\rightarrow 1$, even though the
falloff in $Q^2$ is very rapid for fixed $x$. For specific choices of
$f(x)$ one can thus accomplish $F(Q^2)\sim 1/Q^2$ asymptotically ---
even if $H(x,0,Q^2)$, for $x$ fixed, 
falls off more rapidly with $Q^2$.

In order to illustrate possible consequences of this interplay
between $x$ and $Q^2$ dependences we will  discuss 
the specific class
of exponential models. However, we  emphasize that the 
conclusions 
are not tied to the specific functional dependence chosen,
but  are obtained for 
 all models in which  $H(x,0,Q^2)$ falls faster than $1/Q^2$ 
at fixed $x$, with  
$F(Q^2)\sim 1/Q^2$ from quarks with $x\rightarrow1 $.

Integrating $H(x,0,Q^2)$ of Eq.~(\ref{eq:hx}) over $x$ 
 yields a power law behavior for the form factor
\bea
F(Q^2) \sim \left(\frac{1}{Q^2}\right)^{\frac{m}{n}}.
\eea
However, $m$ and $n$ cannot be arbitrary. In order for the pion to
have a finite size [i.e. integral in Eq. (\ref{eq:master})
converges] , we must have
\bea
(m-2)/n >0 
\eea
Consistency with the observed pion form factor ($F\sim \frac{1}{Q^2}$)
requires $m=n$. Therefore we consider
\bea
H(x,0,Q^2) \stackrel{x\rightarrow 1}
{\longrightarrow} (1-x)^{n-1} \exp \left( -a(1-x)^nQ^2\right),
\quad \quad \quad \left(n>2\right) .
\label{eq:exp}\eea
Straightforward application of Eq. (\ref{eq:master}) yields
\bea
b^2(Q^2) F(Q^2) \sim \int dx (1-x)^{2n-3}
\exp \left( -a(1-x)^nQ^2\right) \sim
\left(\frac{1}{Q^2}\right)^{\frac{2n-2}{n}}.
\eea
Together with $F(Q^2)\sim 1/Q^2$ we thus obtain
\bea
b^2(Q^2)  \sim
\left(\frac{1}{Q^2}\right)^{1-\frac{2}{n}},
\label{eq:b2pi}
\eea
i.e. color transparency occurs only for $n> 2$. The case $n<2$
again corresponds to a situation where the effective
size of the pion increases with $Q^2$
For $n=2$ the effective size of the pion approaches a finite
limit for $Q^2\rightarrow \infty$. In both cases
($n<2$ and $n=2$) there is no color transparency. 

Quark counting rules  predict $q(x)\sim (1-x)^{n-1}$
with $n=2$ for the pion.
This leaves several possibilities:
\begin{enumerate}
\item quark counting rules are correct but the piece in
$H(x,0,Q^2)$ that describes the large $Q^2$ behavior falls
off faster at $x\rightarrow 1$ than $(1-x)$
[i.e. $n>2$]. In this case we should observe color transparency
under suitable experimental conditions.
\item quark counting rules are correct and the same term 
in $H(x,0,Q^2)$ that describes the large $Q^2$ behavior also
describes the $x\rightarrow 1$ behavior. In this case, 
$n=2$ in Eq. (\ref{eq:b2pi}) and there would be  no color transparency.
\item quark counting rules for the PDF are violated and the 
quark distribution
function of the pion vanishes less rapidly than $(1-x)$, i.e.
$n<2$ in Eq. (\ref{eq:exp}). In this case there is no color
transparency --- in fact, in this bizarre case one would even
observe anti-transparency (increasing cross section with $Q^2$).
\end{enumerate}

\subsection{Models Based on Wave Functions}

Factorized models of GPDs do not seem to arise naturally 
from simple dynamical assumptions\cite{Tiburzi:2001ta,Diehl:2003ny}.
Here we assume a system made of two constituents of mass $m$. 
Then rotational invariance
takes the form that the wave function is a function of 
${k_\perp^2+m^2\over x(1-x)}$, where $k_\perp$ is a relative momentum
and the GPD is given by
\bea
H(x,0,Q^2)={2m\over x(1-x)}\int d^2k_\perp \psi^*(x,{\bfk}_\perp+(1-x){\bfq}_\perp)
\psi(x,{\bfk}_\perp),\eea
or 
\bea
H(x,0,\bfB)={2m\over x(1-x)^3}\psi^2(x,{B^2\over (1-x)^2}).\eea

The color transparency aspect of these kinds of models were discussed for the
spin-less case in\cite{prec}. Here we present the case with spin included.
We use the pionic model of Chung, Coester and Polyzou\cite{Chung:mu}
 as a starting point. In that model the form factor is given by 
\bea
&& F_\pi(Q^2)={1\over 4\pi}\int {dx\over x(1-x)}\int d^2k_\perp
[k_\perp^2+m^2+(1-x)\bfk_\perp\cdot {\bfQ}]\phi({\bfk_\perp'^2}) 
\phi({\bfk_\perp^2}), \\
&&\phi({\bfk_\perp^2}) = u(k^2)/[(\bfk_\perp^2+m^2)/x(1-x)]^{1/4},
\;k^2=(\bfk_\perp^2+m^2)/(4x(1-x))-m^2,\\
&&\bfk'=\bfk+(1-x)\bfQ.\eea
The function $u$ is a solution to the wave equation:
\bea
(4(k^2+m^2)+4 mV)u=M^2 u,\eea
with $M$ as the pion mass. Chung et al found that using
a Gaussian form
\bea
u_G(k^2)=(4/\sqrt{\pi}b^3)^{1/2}\exp{(-k^2/2b^2)},\eea
with $m=0.21$ GeV and $b=0.35 $ GeV gave a good description of the
form factor. We shall also use a power law form\cite{Frederico:ye}
\bea
	u_P(k^2)=1/(k^2+b_p^2)^2\sqrt{32b_p^5/\pi},\eea
with $m=0.21$ GeV and $b_p=0.4$ GeV, that  gives a rather similar form factor.

In the present models the GPD is given by
\bea 
H(x,0,Q^2)={1\over 4\pi x(1-x)}\int d^2k_\perp
[k_\perp^2+m^2+(1-x)\bfk_\perp\cdot {\bfQ}]\phi({\bfk_\perp'^2})  
\phi({\bfk_\perp^2}) \eea

The results are displayed  in Figs. ~1, 2.
There is a clear tendency for the peak in $x$ to move to 
increasing values as
$Q^2$ is increasing. This means the mechanism becomes more like that of 
Feynman.
This trend is also 
seen in the computed values of $b^2(Q^2)$ which increase with
$b^2$ as shown in Figs.3.
\begin{figure}
%\begin{Large}
\unitlength1cm %1cm
%first one\begin{picture}(11,9)(0,10.5)
%\begin{picture}(11,9)(0,-10.5)
\begin{picture}(10,8)(-14,0)
  \includegraphics{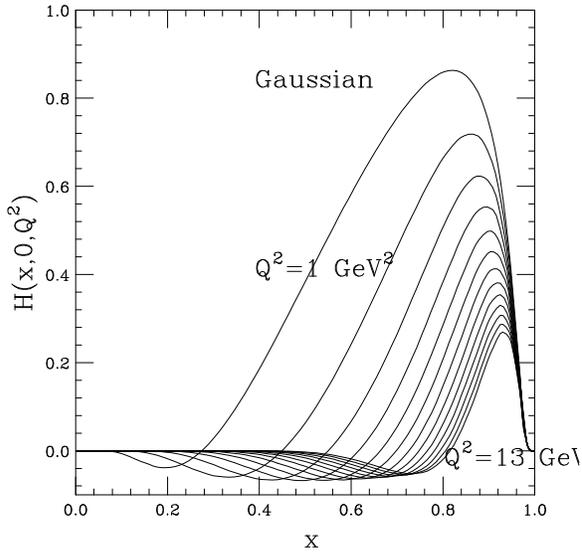}
\end{picture}
%\end{Large}
\label{fig:1}
\caption{GPD for Gaussian model}
\end{figure}
\newpage
\begin{figure}
%\begin{Large}
\unitlength1cm %1cm
%first one\begin{picture}(11,9)(0,-10.5)
%\begin{picture}(11,9)(0,-10.5)
\begin{picture}(10,8)(-15,-0)
  \includegraphics{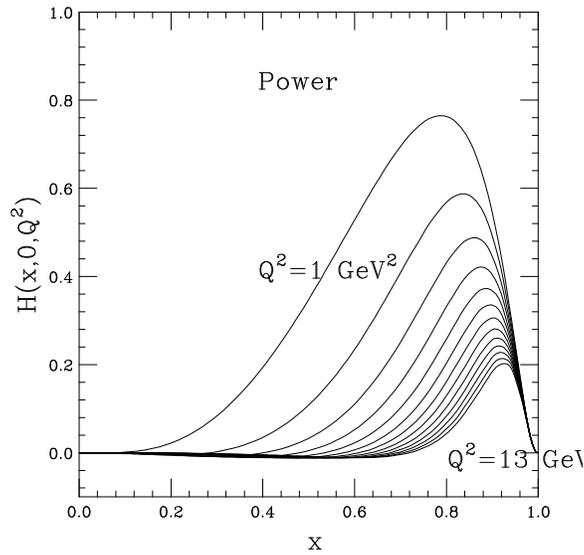}
\end{picture}
%\end{Large}
\label{fig:2}
\caption{GPD for power law  model}
\end{figure}
\begin{figure}
%\begin{Large}
\unitlength1cm %1cm
%first one\begin{picture}(11,9)(0,10.5)
%\begin{picture}(11,9)(0,-10.5)
\begin{picture}(10,8)(-15,0)
  \includegraphics{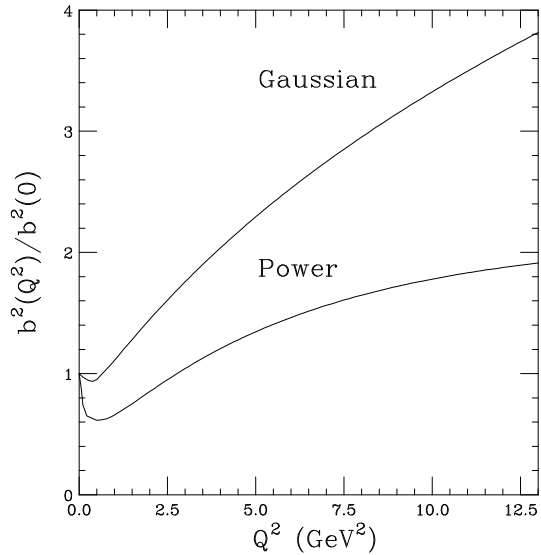}
\end{picture}
%\end{Large}
\label{fig:3}
\caption{$b^2(Q^2)$  for Gaussian and power models}
\end{figure}

These results are obtained for simple models, and lead to conclusions
in sharp contrast with experimental observations\cite{Aitala:2000hb,Aitala:2000hc}. However, they do provide a basis for obtaining a 
 semi-analytic understanding  of $b^2(Q^2)$.
This can be seen by applying the mean
value theorem to the integral over $x$ of Eq.(\ref{eq:master0}).
\bea
b^2(Q^2)F(Q^2)&\approx&
-{\partial \over\partial  Q^2} {1\over (1-\bar{x}(Q^2))^2} F(Q^2)\\
b^2(Q^2)&\approx& {2\over ( 1-\bar{x}(Q^2))^3}{d\bar{x}\over dQ^2}+ {1\over Q^2}.
\label{xbar}
\eea
If $\bar{x}$ is large and increasing then the first term of 
Eq.(\ref{xbar}) will be large and
positive. Then $b^2(Q^2)$ will be large and color transparency would be
precluded. From Figs.~1, 2 
it is clear with that such is the case for those models. With the discovery
of color transparency for pions at our disposal, we can say that these
models are ruled out even though they provide form factors in excellent
agreement with experiment.

In general, 
 precise statements can be made if one knows how $\bar{x}(Q^2)$ approaches
$1$ for $Q^2 \rightarrow \infty$. The crucial question is whether
\bea
\left[1-\bar{x}(Q^2)\right]Q
\eea 
becomes infinite or not as 
$Q^2 \rightarrow \infty$. Let us first consider the case
\bea
\lim_{Q^2\rightarrow \infty}Q \left[1-\bar{x}\right]= 0.
\eea
This happens for example the case when the main $Q^2$ dependence in $H(x,0,Q^2)$
near $x=1$ is through the combination $(1-x)Q^2$. An explicit example is 
provided by a dependence of the form 
$H(x,0,Q^2) \sim q(x)\exp\left[-a(1-x)Q^2\right]$ or 
$H(x,0,Q^2) \sim q(x)\exp\left[-a\frac{(1-x)}{x}Q^2\right]$. In this example,
$\left[1-\bar{x}(Q^2)\right] \sim \frac{1}{Q^2}$ and application of Eq. 
(\ref{xbar}) to $\left[1-\bar{x}(Q^2)\right] \sim \frac{1}{Q^2}$ yields
$b^2(Q^2)\sim Q^2$ and therefore the effective size of the hadron grows with 
increasing $Q^2$.

The opposite happens when 
\bea
\lim_{Q^2\rightarrow \infty}Q \left[1-\bar{x}\right]= \infty.
\eea
An explicit example would be a dependence of the form 
$H(x,0,Q^2) \sim q(x)\exp\left[-a(1-x)^3Q^2\right]$ near $x\rightarrow 1$,
and thus $\left[1-\bar{x}(Q^2)\right] \sim \frac{1}{Q^{2/3}}$. Inserting this 
result into Eq. (\ref{xbar}) yields $b^2(Q^2)\sim Q^{-2/3}$, i.e. the effective
size shrinks with increasing $Q^2$. 

Finally, the marginal case corresponds to
\bea
\lim_{Q^2\rightarrow \infty}Q \left[1-\bar{x}\right]= \mbox{finite}.
\eea
An explicit example is given by 
$H(x,0,Q^2) \sim q(x)\exp\left[-a(1-x)^2Q^2\right]$ near $x\rightarrow 1$,
which yields $\left[1-\bar{x}(Q^2)\right] \sim \frac{1}{Q}$. In this case
\bea
\lim_{Q^2\rightarrow \infty} b^2(Q^2) = const.
\eea

\section{Discussion}

This paper is concerned with the relation between the effective
size of a hadron $b^2(Q^2)$ and generalized parton distributions, 
$H(x,0,Q^2)$. A small effective size is a necessary condition for
color transparency to occur. While a small value of $b^2(Q^2)$ is 
a necessary consequence of perturbative QCD for large enough values
of $Q^2$, the experimental limitation  non-asymptotic 
values of $Q^2$ causes studies of models of strongly interacting QCD
to be 
 relevant and interesting.

Our procedure is to examine a set of model pion wave functions, each giving
rise to a form factor that falls as $1/Q^2$ at large $Q^2$.
This provides a set of  examples  
to illustrate  general principles that control 
the behavior of                   $b^2(Q^2)$ at large $Q^2$.
These  can be summarized as follows
\begin{itemize}
\item If the GPD-representation for the form factor, $H(x,0,Q^2)$,
 is dominated by the average value of $x$ (see Eq.~(\ref{xbar}),
$\bar{x} \neq 1$ at large $Q^2$ (e.g. PQCD mechanism) then the effective size
$b^2(Q^2)$ goes to zero at large $Q^2$. Then  
color transparency is expected to  occur
 under suitable experimental conditions.
\item On the other hand, if the GPD representation of the form factor is
dominated by $\bar{x}\rightarrow 1$ (e.g. Feynman mechanism) then the crucial 
question is how rapidly does $\bar{x}$ approach $1$ with increasing $Q^2$.
\item The effective size of the hadron goes to zero at large $Q^2$ (a necessary
condition for color transparency) only if $\lim_{Q^2\rightarrow \infty}
\left[1-\bar{x}(Q^2)\right]Q=\infty$.
\end{itemize}

The fact that
color transparency has been observed \cite{Aitala:2000hb,Aitala:2000hc}
indicates that the piece in
$H(x,0,Q^2)$ that describes the large $Q^2$ behavior does fall
off faster at $x\rightarrow 1$ than $(1-x)$. In the notation of
Eq.~(\ref{eq:exp}), $n>2$. More generally, we can be confident that
$\lim_{Q^2\rightarrow \infty}
\left[1-\bar{x}(Q^2)\right]Q=\infty$.
 This means that
either its GPD at fixed $x$ falls off like the
form factor, or the GPD which describes the 
leading behavior of the form factor vanishes near $x\rightarrow 1$
faster than $(1-x)$.

 Our results
 provide a  test for GPDs, and also  a
further test for light-cone wave functions of the pion.
The simple pion wave function models used here
are not consistent with the  dependence on $x$
required to obtain a small $b^2(Q^2)$, and we consider these to be
ruled out. But there are a variety of models in the literature\cite{Maris:2000sk},
and we suggest that $b^2(Q^2)$ be evaluated to provide further tests.

 Experimental 
studies of pionic color transparency are planned\cite{exp}. 
If color transparency is
observed at momentum transfers accessible to Jefferson Laboratory,
then our results provide predictions regarding the pionic GPD that
could be tested by (difficult, but not impossible) experiments.

%GPDs provide a decomposition of the form factor with respect to the 
%light-cone momentum of the active quark.
%One can distinguish two general classes of functions that describe
%GPDs. 

%In the most simple case the GPD at fixed $x$ falls off like the
%form factor. In this case 
%$b^2(Q^2) \stackrel{Q^2\rightarrow\infty}{\longrightarrow} 0$ and
%we expect color transparency.

%The more subtle case is the one where the GPDs at $x<1$ fall
%off faster than the form factor and where the form factor at
%large $Q^2$ is supported by quarks carrying momentum fractions
%$x\rightarrow 1$. In this case the result depends on the details
%of the behavior of $H(x,0,Q^2)$ near $x\rightarrow 1$.

%We assumed a functional form where the GPD near $x\rightarrow 1$
%is a rapidly falling function of $(1-x)^n Q^2$ 
%times some power of $(1-x)$. Such an ansatz can still yield the 
%desired power law behavior of the form factor, provided the powers of 
%$(1-x)$ are selected appropriately.
%In this case, if both the form factor at large $Q^2$ and the
%leading behavior of the PDF near $x\rightarrow 1$ are described by
%the same term, and if quark counting rules are satisfied then
%color transparency cannot occur. However, color transparency can
%still occur provided that the term in the GPD which describes the 
%leading behavior of the form factor vanishes near $x\rightarrow 1$
%faster than predicted by quark counting rules [i.e. faster than
%$(1-x)$ for the pion and faster than $(1-x)^3$ for the nucleon].

{\bf Acknowledgements:}
We would like to thank M. Diehl for useful dicussions. This work
was supported by the DOE under contracts
DE-FG03-96ER40965 (MB) and DE-FG02-97ER41014 (GM).


\begin{thebibliography}{99}
\bibitem{mueller} A.H.~Mueller in Proceedings of Seventeenth Rencontre de
    Moriond, Moriond, 1982 ed. J Tran Thanh Van (Editions Frontieres,
Gif-sur-Yvette, France, 1982)p13.


\bibitem{brodsky}S.J.~Brodsky in Proceedings of the Thirteenth International Symposium
on Multiparticle Dynamics, ed. W.~Kittel, W.~Metzger and A.~Stergiou (World
Scientific, Singapore 1982,) p963.

%\cite{Frankfurt:1992dx}
\bibitem{Frankfurt:1992dx}
L.~Frankfurt, G.~A.~Miller and M.~Strikman,
%``Color transparency phenomenon and nuclear physics,''
Comments Nucl.\ Part.\ Phys.\  {\bf 21}, 1 (1992).
%%CITATION = CNPPA,21,1;%%

\bibitem{FMS94}
  L.L. Frankfurt, G.A. Miller and M. Strikman, 
Ann. Rev. Nucl. Part. Sci. {\bf 44}, 501 (1994).
%%CITATION = HEP-PH 9407274;%%

%\cite{Jain:1995dd}
\bibitem{Jain:1995dd}
P.~Jain, B.~Pire and J.~P.~Ralston,
%``Quantum Color Transparency and Nuclear Filtering,''
Phys.\ Rept.\  {\bf 271}, 67 (1996)
[arXiv:hep-ph/9511333].
%%CITATION = HEP-PH 9511333;%%

%\cite{Jennings:1989hc}
\bibitem{Jennings:1989hc}
B.~K.~Jennings and G.~A.~Miller,
%``On Color Transparency,''
Phys.\ Lett.\ B {\bf 236}, 209 (1990);
%%CITATION = PHLTA,B236,209;%%
B.~K.~Jennings and G.~A.~Miller,
%``Color Transparency: The Wherefore And The Why,''
Phys.\ Rev.\ D {\bf 44}, 692 (1991).
%%CITATION = PHRVA,D44,692;%%

\bibitem{prec} 
L.~Frankfurt, G.~A.~Miller and M.~Strikman,
%``Precocious Dominance Of Point - Like Configurations In Hadronic Form-Factors,''
Nucl.\ Phys.\ A {\bf 555} (1993) 752.
%%CITATION = NUPHA,A555,752;%%
%\cite{Low:1975sv}
\bibitem{Low:1975sv}
F.~E.~Low,
%``A Model Of The Bare Pomeron,''
Phys.\ Rev.\ D {\bf 12}, 163 (1975).
%%CITATION = PHRVA,D12,163;%%

%\cite{Gunion:iy}
\bibitem{Gunion:iy}
J.~F.~Gunion and D.~E.~Soper,
%``Quark Counting And Hadron Size Effects For Total Cross-Sections,''
Phys.\ Rev.\ D {\bf 15}, 2617 (1977).
%%CITATION = PHRVA,D15,2617;%%

%\cite{Blaettel:rd}
\bibitem{Blaettel:rd}
B.~Blaettel, G.~Baym, L.~L.~Frankfurt and M.~Strikman,
%``How Transparent Are Hadrons To Pions?,''
Phys.\ Rev.\ Lett.\  {\bf 70}, 896 (1993).
%%CITATION = PRLTA,70,896;%%

%\cite{Frankfurt:1996ri}
\bibitem{Frankfurt:1996ri}
L.~Frankfurt, A.~Radyushkin and M.~Strikman,
%``Interaction of small size wave packet with hadron target,''
Phys.\ Rev.\ D {\bf 55}, 98 (1997)
[arXiv:hep-ph/9610274].
%%CITATION = HEP-PH 9610274;%%

\bibitem{Aitala:2000hb}
E.~M.~Aitala {\it et al.}  [E791 Collaboration],
%``Direct measurement of the pion valence quark momentum distribution, the  pion light-cone wave function squared,''
Phys.\ Rev.\ Lett.\  {\bf 86}, 4768 (2001)
[arXiv:hep-ex/0010043].
%%CITATION = HEP-EX 0010043;%%
\bibitem{Aitala:2000hc}
E.~M.~Aitala {\it et al.}  [E791 Collaboration],
%``Observation of color-transparency in diffractive dissociation of pions,''
Phys.\ Rev.\ Lett.\  {\bf 86}, 4773 (2001)
[arXiv:hep-ex/0010044].
%%CITATION = HEP-EX 0010044;%%
\bibitem{Frankfurt:2000jm}
L.~Frankfurt, G.~A.~Miller and M.~Strikman,
 %``Coherent QCD phenomena in the coherent pion nucleon and pion nucleus
%production of two jets at high relative momenta,''
Phys.\ Rev.\ D {\bf 65}, 094015 (2002)
[arXiv:hep-ph/0010297].
%%CITATION = HEP-PH 0010297;%%

\bibitem{Frankfurt:it}
L.~Frankfurt, G.~A.~Miller and M.~Strikman,
%``Coherent Nuclear Diffractive Production Of Mini - Jets: Illuminating Color Transparency,''
Phys.\ Lett.\ B {\bf 304}, 1 (1993)
[arXiv:hep-ph/9305228].
%%CITATION = HEP-PH 9305228;%%




\bibitem{mb1} M. Burkardt, Phys.\ Rev.\ D {\bf 62}, 071503 (2000),
Erratum-ibid. D {\bf 66}, 119903 (2002);
M.~Burkardt,
%``Impact parameter space interpretation 
%for generalized parton  distributions,''
Int.\ J.\ Mod.\ Phys.\ A {\bf 18}, 173 (2003)
%[arXiv:hep-ph/0207047].
%%CITATION = HEP-PH 0207047;%%

\bibitem{soper} D.E. Soper, Phys.\ Rev.\ D\ {\bf 15}, 1141 (1977).

\bibitem{pos} M. Burkardt, in `Lepton Scattering, Hadrons and QCD',
Eds.: W. Melnitchouk et al., Adelaide, Australia, March 2001,
hep-ph/0105324; P.V. Pobylitsa, Phys.\ Rev.\ D\ {\bf 66}, 094002 
(2002).

\bibitem{Radyushkin:1998rt}
A.~V.~Radyushkin,
%``Nonforward parton densities and soft mechanism for form factors and  wide-angle Compton scattering in {QCD},''
Phys.\ Rev.\ D {\bf 58}, 114008 (1998)
[arXiv:hep-ph/9803316].
%%CITATION = HEP-PH 9803316;%%

                     
\bibitem{Tiburzi:2001ta}
B.~C.~Tiburzi and G.~A.~Miller,
%``Exploring skewed parton distributions with two-body models on the light  front. I: Bimodality,''
Phys.\ Rev.\ C {\bf 64}, 065204 (2001)
[arXiv:hep-ph/0104198].
%%CITATION = HEP-PH 0104198;%%
\bibitem{Diehl:2003ny}
M.~Diehl,
%``Generalized parton distributions,''
Phys.\ Rept.\  {\bf 388}, 41 (2003)
[arXiv:hep-ph/0307382].
%%CITATION = HEP-PH 0307382;%%
\bibitem{Chung:mu}
P.~L.~Chung, F.~Coester and W.~N.~Polyzou,
%``Charge Form-Factors Of Quark Model Pions,''
Phys.\ Lett.\ B {\bf 205}, 545 (1988).
%%CITATION = PHLTA,B205,545;%%

%\cite{Frederico:ye}
\bibitem{Frederico:ye}
T.~Frederico and G.~A.~Miller,
%``Null Plane Phenomenology For The Pion Decay Constant And Radius,''
Phys.\ Rev.\ D {\bf 45} (1992) 4207.
%%CITATION = PHRVA,D45,4207;%%

%\cite{Maris:2000sk}
\bibitem{Maris:2000sk}
P.~Maris and P.~C.~Tandy,
%``The pi, K+, and K0 electromagnetic form factors,''
Phys.\ Rev.\ C {\bf 62}, 055204 (2000)
[arXiv:nucl-th/0005015];
%%CITATION = NUCL-TH 0005015;%%%\cite{Nesterenko:1982gc}
V.~A.~Nesterenko and A.~V.~Radyushkin,
%``Sum Rules And Pion Form-Factor In QCD,''
Phys.\ Lett.\ B {\bf 115}, 410 (1982);
%%CITATION = PHLTA,B115,410;%%
F.~Cardarelli, E.~Pace, G.~Salme and S.~Simula,
%``Nucleon and pion electromagnetic form-factors in a light front constituent quark model,''
Phys.\ Lett.\ B {\bf 357}, 267 (1995)
[arXiv:nucl-th/9507037];
%%CITATION = NUCL-TH 9507037;%%%\cite{Stefanis:1998dg}
N.~G.~Stefanis, W.~Schroers and H.~C.~Kim,
%``Pion form factors with improved infrared factorization,''
Phys.\ Lett.\ B {\bf 449}, 299 (1999)
[arXiv:hep-ph/9807298];
%%CITATION = HEP-PH 9807298;%%
W.~Broniowski and E.~Ruiz Arriola,
%``Impact-parameter dependence of the generalized parton distribution of the pion in chiral quark models,''
Phys.\ Lett.\ B {\bf 574}, 57 (2003)
[arXiv:hep-ph/0307198];
%%CITATION = HEP-PH 0307198;%%
C.~D.~Roberts and A.~G.~Williams,
%``Dyson-Schwinger Equations And Their Application To Hadronic Physics,''
Prog.\ Part.\ Nucl.\ Phys.\  {\bf 33}, 477 (1994)
[arXiv:hep-ph/9403224].
%%CITATION = HEP-PH 9403224;%%




\bibitem{exp} Jefferson Laboratory E-01-107, K.~R.~ Garrow and R. Ent spokespersons

\end{thebibliography}
\end{document}